\documentclass[5p,times]{elsarticle}
\usepackage{epsfig}
\usepackage{graphicx}
\usepackage{amssymb}
\usepackage[figuresright]{rotating}
\usepackage{amssymb}
\begin{document}
\begin{frontmatter}
\title{Two-component model in quantum statistical framework compared with multiplicity distributions in proton-proton collisions at energies up to $\sqrt {s}$ = 7 TeV}

\author{\it Premomoy Ghosh}
\ead{prem@vecc.gov.in}
\address{Variable Energy Cyclotron 
Centre, 1/AF Bidhan Nagar, Kolkata 700 064, India}            

\begin{abstract}
Proton-proton collisions at new high energies ($\sqrt {s} =$ 2.36 and 7 TeV) at LHC resulted 
into greater mean multiplicities ($\langle n \rangle$) of charged particles in the 
mid-rapidity region than estimated ones by different models and event generators. 
Another significant observation in multiplicity data is the change in slope in the distribution 
of primary charged hadrons in symmetric pseudorapidity interval $|\eta|<$2.4. The change is most prominent 
with data at $\sqrt{s} = 7$ TeV. These new observations merit further studies. We consider a two-component model of 
particle production to analyze multiplicity distributions of charged hadrons from proton-proton collisions at 
centre-of-mass energies $\sqrt{s} = $ 0.9, 2.36 and 7 TeV in symmetric pseudorapidity intervals $|\eta|$ of 
increasing width around the centre-of-mass pseudorapidity $\eta_{cm} = 0$. The model, based on 
quantum statistical (QS) formalism, describes multiplicity distribution by convolution of a Negative 
Binomial Distribution (NBD), representing a chaotic component, and a Poisson Distribution (PD), 
representing a coherent component of particle productions. The behaviour of characteristic 
parameters of the model is followed by the LHC data, while a scaling law, involving information entropy 
in quantum statistical viewpoint and derived as a function of chaotic multiplicity obtained from the 
two-component model, is not obeyed by the data, satisfactorily. An attempt to match the measured multiplicity
distributions and suggested convolutions with values of characteristic parameters extracted from the data
confirms disagreement between the data and the model.
\end{abstract}

\end{frontmatter}
\section{Introduction}
\label{}
The probability distribution $P_{n}(n)$ of production of $n$ particles from 
collisions, has been the premier of basic observables, characterizing the 
final states of multi-particle production process in high energy physics 
experiments since the beginning of such studies. In recent times, the Large
Hadron Collider (LHC) \cite{ref01} at CERN has taken high energy collisions
in laboratories to a new energy domain, facilitating proton-proton ($pp$) 
collisions at centre-of-mass energies, $\sqrt {s}$ = 0.9, 2.36 and 7 TeV 
\cite{ref02,ref03,ref04,ref05}. The Compact Muon Solenoid (CMS) experiment 
at LHC has measured multiplicity distributions of primary charged hadrons for all the 
three LHC energies, available so far, in the mid-pseudorapidity ($\eta$) region 
(where $\eta = -ln[tan(\theta/2)]$ and  $\theta$ is the polar angle of the particle 
with respect to the counterclockwise beam direction) in five symmetric overlapping 
$\eta$-intervals $|\eta|$ or $\eta_{c}<$ 0.5, 1.0, 1.5, 2.0 and 2.4 \cite{ref04} around the 
centre-of-mass pseudorapidity ($\eta_{cm}=0$). The measured mean multiplicities at the new LHC 
energies ($\sqrt {s} =$2.36 and 7 TeV) in the range of mid-pseudorapidity, 
have been found to be largely underestimated by existing models (event generators 
like PYTHIA, PHOJET etc). Other experiments at LHC corroborate the finding, generating
strong motivation to look into the LHC data from different approaches. Another 
significant observation by the CMS experiment is the change in slope in the distribution 
of primary charged hadrons in pseudorapidity interval $|\eta|<2.4$ at $\sqrt{s} = 7$ TeV. 
Appearance of a change in slope in the measured distribution indicates to the possible existence of 
more-than-one process or source of particle production. In this article, we present results of our analysis 
of multiplicity distributions data from $pp$ collisions at LHC, 
in the light of a two-component model \cite{ref06} formulated in a quantum statistical 
approach. We extend our study further to check the validity of a scaling law involving information 
entropy in quantum statistical point of view. 
\section{Background}
\label{}
In the last few decades, with the advent of accelerator technology, collider 
facilities, capable of delivering higher and higher centre-of-mass energy 
($\sqrt{s}$), could be made available for proton-proton ($pp$) and proton-
antiproton ($p\bar p$) collisions. While the Intersecting Storage Ring (ISR), 
CERN facilitated pp collisions at $\sqrt{s}$ range of 23.9 to 62.2 GeV \cite{ref07}, 
the Super Proton Synchrotron (SPS), CERN had $p\bar p$ collisions at $\sqrt{s} = 200$ to 
$\sqrt{s} = 900$ GeV \cite{ref08}. In the Tevatron at Fermi lab, the energies 
of $p\bar p$ collisions were $\sqrt{s} = 540$ GeV to $\sqrt {s} = 1.8$ TeV \cite{ref09} 
and finally at the Large Hadron Collider (LHC), CERN, energy of collisions of protons have 
reached $\sqrt{s}$ as high as 7 TeV \cite{ref03,ref04,ref05}. With all these data
along with data at the fixed-target (non-collider) experiments at pre-ISR 
period, a rich set of experimental data on multiplicity distributions from 
$pp$ ($p\bar p$) collisions is now available for a wide range of $\sqrt {s}$ for a 
comprehensive and systematic study in different theoretical and phenomenological 
formalisms for better understanding of multiparticle production mechanism. 
The two-component model \cite{ref06} of particle productions is one such 
formalism which has been thoroughly used in analyzing experimental data of 
$pp$ ($p\bar p$) collisions at energies available up to the SPS.

Of the statistical distribution functions, the Negative Binomial Distribution (NBD),
\begin{equation}
  P(n,\langle n \rangle, k) = \frac{(n+k-1)!}{n!(k+1)!}\left[\frac{\langle n \rangle}{k+\langle n \rangle}\right]^n \times \left[\frac{k}{k+\langle n \rangle}\right]^k
\end{equation}  
has been the most successful one in describing the probability 
distributions of final state charged particles from $pp$ ($p\bar p$) collisions
in the discussed energy domain. At the lower part of the energy range, 
the multiplicity distributions of final state charged particles in the full 
phase space could be described by binomial distribution (when the parameter
$k$ in equation-1 is negative and an integer). At around $\sqrt {s}$ = 5 GeV, the 
distribution of produced charged particles turned broader and started 
following Poisson distribution ($k$ is infinite). Above $\sqrt {s}$ = 30 GeV, the 
NBD matched fairly well with even broader distributions in the full phase 
space (pseudorapidity space) data up to $\sqrt {s} =$ 540 GeV at SPS. At 
$\sqrt{s} = 900$ GeV at SPS, when a shoulder-like structure appeared in the 
tail of the multiplicity distributions, a single NBD failed in matching with 
the data at large pseudorapidity intervals. At 900 GeV and also at 1.8 TeV at 
Tevatron, the NBD function was successful only in the restricted mid-pseudorapidity 
($|\eta |<0.5$) region. The distribution in large $\eta$-interval could be reasonably 
described with the sum of two NBDs.
\section{Motivation}
\label{}
At the LHC energies, multiplicity distributions in non-single diffractive 
(NSD) inelastic proton-proton collisions have been measured and reported 
by different experiments, in different kinematic ranges, depending on
capability of respective detector setup in terms of geometrical acceptance, 
detection efficiencies etc. A Large Ion Collider Experiment (ALICE) at LHC has 
measured primary charged particles at $\sqrt {s}$ = 0.9 and 2.36 TeV in the mid-$\eta$ 
region in three overlapping $\eta$-intervals $|\eta|<$  0.5, 1.0 and 1.3 \cite{ref02}. 
At $\sqrt {s}$ = 7 TeV, instead of NSD inelastic events, ALICE analyzed \cite{ref03}  
an event class requiring at least one charged particle in $|\eta |<1$ and measured 
multiplicity distribution in that $\eta$-interval only. The ATLAS experiment at LHC has 
measured charged particle multiplicities for different event classes characterized 
by different lower cuts on the number of charged particles ($n_{ch} < 1$, 2 and 6) 
in different kinematic ranges ($p_{T}>$ 100 MeV, 500 MeV in $|\eta|<$2.5). The distributions 
measured by ALICE in the three $|\eta |$-intervals, $|\eta|<$ 0.5, 1.0 and 1.3, at 
the two energies, $\sqrt{s} = 0.9$ TeV, $\sqrt{s} = 2.36$ TeV have been reported 
to match fairly well with NBD. The NBD fit to the distribution at $\sqrt {s}$ = 7 TeV, 
measured by ALICE, has been reported to be slightly underestimating the data at 
low multiplicity ($n <5$) and slightly overestimating the data at high multiplicity 
($n >55$). Measurements of multiplicity distributions in a wider phase space and in all 
the three energies by CMS experiment, reveal interesting features. Presenting \cite{ref04} 
multiplicity distributions, without fitting to any distribution function, CMS 
reported a change of slope in $P_{n}$ for $n > 20$ in its largest $\eta$-interval 
of $|\eta|<2.4$. This feature becomes more pronounced with increasing $\sqrt{s}$. 

The observed change in slope in the distributions as measured 
by CMS could be attributed to the existence of more than one kind of source 
or process of particle production. There have been several proposals of 
models \cite{ref06,ref10,ref11,ref12,ref13} involving multi-particle production from 
more than one process / source. In Ref.\cite{ref10}, it is the weighted superposition of 
two multiplicity distributions (each assumed to follow NBD); one due to soft events (without 
mini-jets) and the other representing semi-hard events (with mini-jets). In the multiparton 
interactions model described in Refs.\cite{ref11, ref12}, the soft component, which constitute the 
bulk of events in the final state, corresponds to single parton-parton collisions in the framework of 
dual parton model and produces KNO \cite{ref14} distribution. The other process, which seems to be 
superimposed on top of the KNO producing process, involves two or more independent parton-parton collisions. 
Ref.\cite{ref13} is a two-component dual parton model (DPM), the soft process is described by the 
supercritical Pomeron and the hard component is described by perturbative QCD. This two-component DPM 
includes diffractive processes also. In spite of all these efforts, the change in slope in distributions 
is not yet a fully understood phenomenon. 

We recollect that the distribution at $\sqrt {s} =$ 900 GeV in broad phase space has been explained 
by sum of two NBDs. Also, there has been agreement \cite{ref12} on the soft parts (as measured in UA5 and 
E735 experiments) obtained by the models described in Ref.\cite{ref10} and Ref.\cite{ref11}. 
But, at this point, it is important to consider relative position of the change in slope in the 
distributions of data at $\sqrt {s} = $ 900 GeV at SPS and $\sqrt {s} = $ 7 TeV at LHC. The change 
in slope appeared in SPS data at a higher multiplicity ($n$) and a sum of two NBDs (representing two 
broad distributions) could describe the data satisfactorily. For the LHC data, the most prominent change in 
slope appears in the multiplicity distribution  in the symmetric pseudorapidity interval, $\eta_{c}<$ 
2.4 at $\sqrt {s} =$7 TeV at much lower multiplicity ($n \rangle 20$), indicating that the measured distribution 
could be due to contributions from a broad distribution and a narrow distribution. 

Such a scenario is provided by a two-component model \cite{ref06,ref15,ref16, ref17} in quantum statistical  approach, 
where a weighted average of a NBD (representing  a broad distribution) with $k=1$ due to a chaotic source 
and a Poisson distribution (representing a narrow distribution) due to a coherent source results into the 
final distribution. We follow this two-component model in QS viewpoint to analyze the LHC data.
\section{Methodology}
\label{}
In the two-component model \cite{ref06} in QS approach, the total mean multiplicity is given by:
\begin{equation}
\langle n \rangle = \langle n_{ch}\rangle + \langle n_{co}\rangle  
\end{equation} 
where $\langle n_{co} \rangle$ is the coherent component and 
$\langle n_{ch} \rangle$ is the chaotic component of the mean of total 
multiplicity.

In quantum statistics, a completely chaotic source produces particles following
a negative binomial distribution given by equation-1 ($n$ in equation-1 is to 
be read as $n_{ch}$ in the present formalism), where k is the number of 
cells in the phase space or the number of independent quantum states. On the 
other hand, a completely coherent source gives rise to a Poisson distribution: 
      
\begin{equation}
P(\langle n_{co} \rangle, k_{p}) = \frac{\langle n_{co} \rangle^{k_{p}} e^{-\langle n_{co} \rangle}}{k_{p}!}
\end{equation}

The mean of the chaotic component of total multiplicity, $\langle n_{ch} 
\rangle$ is obtained from the measured mean of total multiplicity and the 
second moment of the distribution by:
\begin{equation}
\langle n_{ch} \rangle = \tilde{p} \langle n \rangle 
\end{equation}
where
\begin{equation}
\tilde{p} = [K \{C_2 - (1+1/\langle n \rangle) \}]^{1/2}
\end{equation}
is the chaoticity parameter, $C_2 = \langle n^2 \rangle/\langle n \rangle^2$ is 
the second moment of multiplicity distribution and $K$ can have value either 1 
or 2. However, in narrow symmetrical $\eta$ intervals in data, only $K$=1
leads to physical solutions, in most of the cases. The chaoticity, 
$\tilde{p}$ ($0<\tilde{p}<1$) indicates measure of fraction of chaoticity 
involved in the source. For a completely chaotic source, $\tilde{p} = 1$ 
and for a completely coherent source, $\tilde{p} = 0$. Any value, in between, 
represents a convoluted distribution of the two. 

For our present study, we use the published \cite{ref04} multiplicity distribution 
data from CMS experiments in five overlapping symmetric pseudorapidity intervals 
$|\eta|$ around the centre-of-mass pseudorapidity with width of the interval extended up to 
$|\eta|<$ 2.4 for the three LHC energies, $\sqrt {s}$ = 0.9, 2.36 and 7 TeV. In some cases, we use 
published ALICE \cite{ref02} data in $|\eta|<$ 0.5, 1.0 and 1.3 for $\sqrt {s}$ = 0.9 and 
2.36 TeV for which the  multiplicity distributions in non-single diffractive (NSD) 
inelastic proton-proton collisions have been measured and published by the experiment. 
\section{Results and Discussions}
\label{}
\subsection{Behaviour of characteristic parameters of the model}
\label{}
\begin{center}
\begin{figure}[htbp]
\includegraphics[scale=0.44]{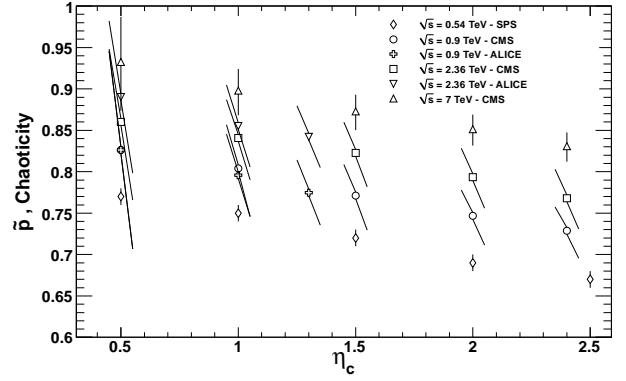}
 \caption{Chaoticity for $\sqrt {s}$ = 0.54, 0.9, 2.36 and 7.0 TeV for different $\eta$-intervals. The error-bars, include both the statistical and the systematic uncertainties. Some of the error-bars are drawn inclined for clarity in presentation.}
\label{fig:chaoticity} 
\end{figure}
\end{center}

\begin{center}
\begin{figure}[htbp]
\includegraphics[scale=0.44]{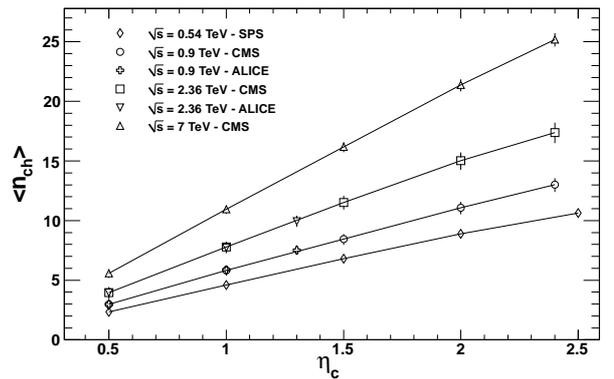}
\caption{Mean of chaotic multiplicity for $\sqrt {s}$ = 0.54, 0.9, 2.36 and 7.0 TeV for different $\eta$-intervals. Lines drawn by joining data-points to guide the eye. The error-bars include both the statistical and the systematic uncertainties.}
\label{fig:chaotic_multi} 
\end{figure}
\end{center}

Following Fowler et al.\cite{ref06}, using equations-5 and 4 given in section-4 above, 
we calculate chaoticity ($\tilde{p}$) and mean of the chaotic multiplicity ($\langle n_{ch} \rangle$) 
for all the multiplicity distributions in NSD $pp$ collisions as reported\cite{ref02,ref04} by the two 
LHC experiments, CMS and ALICE. We plot the parameters in Fig.~\ref{fig:chaoticity} 
and Fig.~\ref{fig:chaotic_multi} respectively. In these plots, we include similar data for 
$\sqrt {s} = 540$ GeV of SPS also for comparison. The chaoticity depicts fractional contribution 
of chaotic source in the total multiplicity. As could be seen in Fig.~\ref{fig:chaoticity}, the chaoticity 
decreases with increasing width of the $\eta$-interval, $\eta_{c}$ for a given $\sqrt {s}$. Also, for a given 
width of $\eta_{c}$, the chaoticity increases with $\sqrt {s}$. 
It is worth mentioning at this point that, 
for the LHC data, the statistical and the systematic uncertainties as quoted with probability 
distribution data \cite{ref02,ref04} have been added in quadrature and propagated all through 
the analysis. The error-bars associated with the LHC-data points in the plots in all the figures of
this article, therefore, include both the statistical and the systematic uncertainties. The plots of chaotic 
multiplicity in Fig.~\ref{fig:chaotic_multi} shows rapid increase in chaotic multiplicity with both the 
$\eta_{c}$ and $\sqrt {s}$. 

\begin{center}
\begin{figure}[htbp]
\includegraphics[scale=0.44]{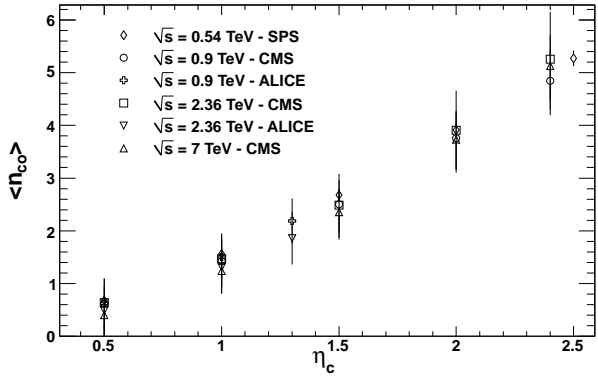}
\caption{Mean of coherent multiplicity for $\sqrt {s}$ = 0.54 TeV, 0.9 TeV, 2.36 TeV and 7.0 TeV for different $\eta$-intervals. The error-bars include both the statistical and the systematic uncertainties.}
\label{fig:coherent_multi} 
\end{figure}
\end{center}

We also calculate the coherent component ($\langle n_{co} \rangle$) of mean multiplicity using equation-2 in section-4 
and study the dependence of $\langle n_{co} \rangle$ on $\eta_{c}$ and $\sqrt {s}$. Our analysis 
reveals (as has been shown in Fig.~\ref{fig:coherent_multi}) almost no change in $\langle n_{co} \rangle$  
with increasing $\sqrt {s}$ in a given $\eta_{c}$ and a very low rate of increase in $\langle n_{co} \rangle$
(as compared to the rate of increase in $\langle n_{ch} \rangle$) with increase in $\eta_{c}$ for a given $\sqrt {s}$, 
in the considered mid-pseudorapidity region. 
Comparing observed dependences of the parameters of the model for the LHC data on $\eta_{c}$ and $\sqrt {s}$,
the particle production scenario at the considered energy range by the present approach could be summerized
as follows:

a) The chaotic source populates mainly the mid-pseudorapidity region and the 
fractional contribution of the chaotic source as compared to the coherent source 
decreases as one goes away from the centre-of-mass pseudorapidity ($\eta_{cm}=0$). 

b) The span of pseudorapidity of dominance of chaotic source increases with 
increase in $\sqrt {s}$.

c) The increase in the total mean multiplicity $\langle n \rangle$ with $\sqrt {s}$ in the mid-pseudorapidity region, in the considered 
energy range, is dominantly due to the increase in chaotic multiplicity $\langle n_{ch} \rangle$ with $\sqrt {s}$.

In general, the behaviour of the parameters extracted from the LHC data match with characteristic features 
of the model in a similar way as has been observed in case of SPS energy data. It is naturally logical to analyze
the LHC data in terms of a scaling law \cite{ref18}, involving chaotic multiplicity, which has been reported to 
hold for $pp$ or $p\bar p$ data for a wide range of energies at ISR and SPS.   
\subsection{Scaling of  information entropy}
\label{}
\begin{center}
\begin{figure}[htbp]
\includegraphics[scale=0.44]{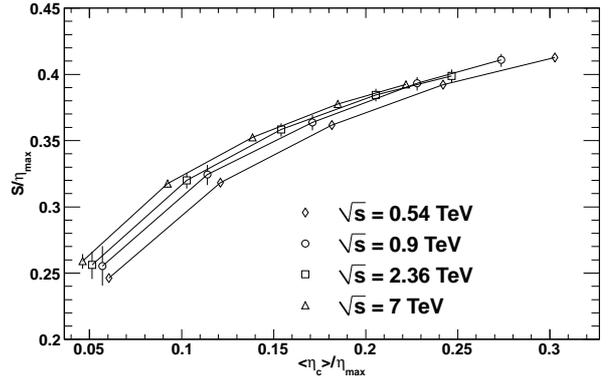}
\caption{The information entropy, calculated with parameters of the 
two-component model in QS approach, as extracted from the LHC data do
not appear to obey the scaling law \cite{ref18}.}
\label{fig:entropy_scaling} 
\end{figure}
\end{center}

We analyze the LHC data of NSD $pp$ collisions up to $\sqrt {s}$ = 7 TeV, as measured by the CMS experiment to 
test the validity of a proposed scaling law namely the scaling of information entropy \cite{ref18} involving 
application of the discussed two component model. The information entropy is a function of the chaotic multiplicity 
($<n_{ch}>$), which is a function of symmetric pseudorapidity interval, $\eta_{c}$ and the centre-of-mass energy, 
$\sqrt {s}$. The entropy for chaotic multiplicities in symmetric pseudorapidity intervals $\eta_{c}$ is given by:
\begin{eqnarray}
\nonumber S(\eta_{c},\sqrt{s})
\nonumber &=& (<n_{ch}>+1)ln(<n_{ch}>+1)\\
 &-& <n_{ch}> ln <n_{ch}>
\end{eqnarray}
In Fig.~\ref{fig:entropy_scaling}, we plot $S/\eta_{max}$ as a function of $\xi = \eta_{c}/\eta_{max}$ 
where \cite{ref18},
\begin{equation}
\eta_{max}=ln[(\sqrt{s} - 2m_n)/m_\pi]
\end{equation} 
\begin{center}
\begin{figure}[htbp]
\includegraphics[scale=0.44]{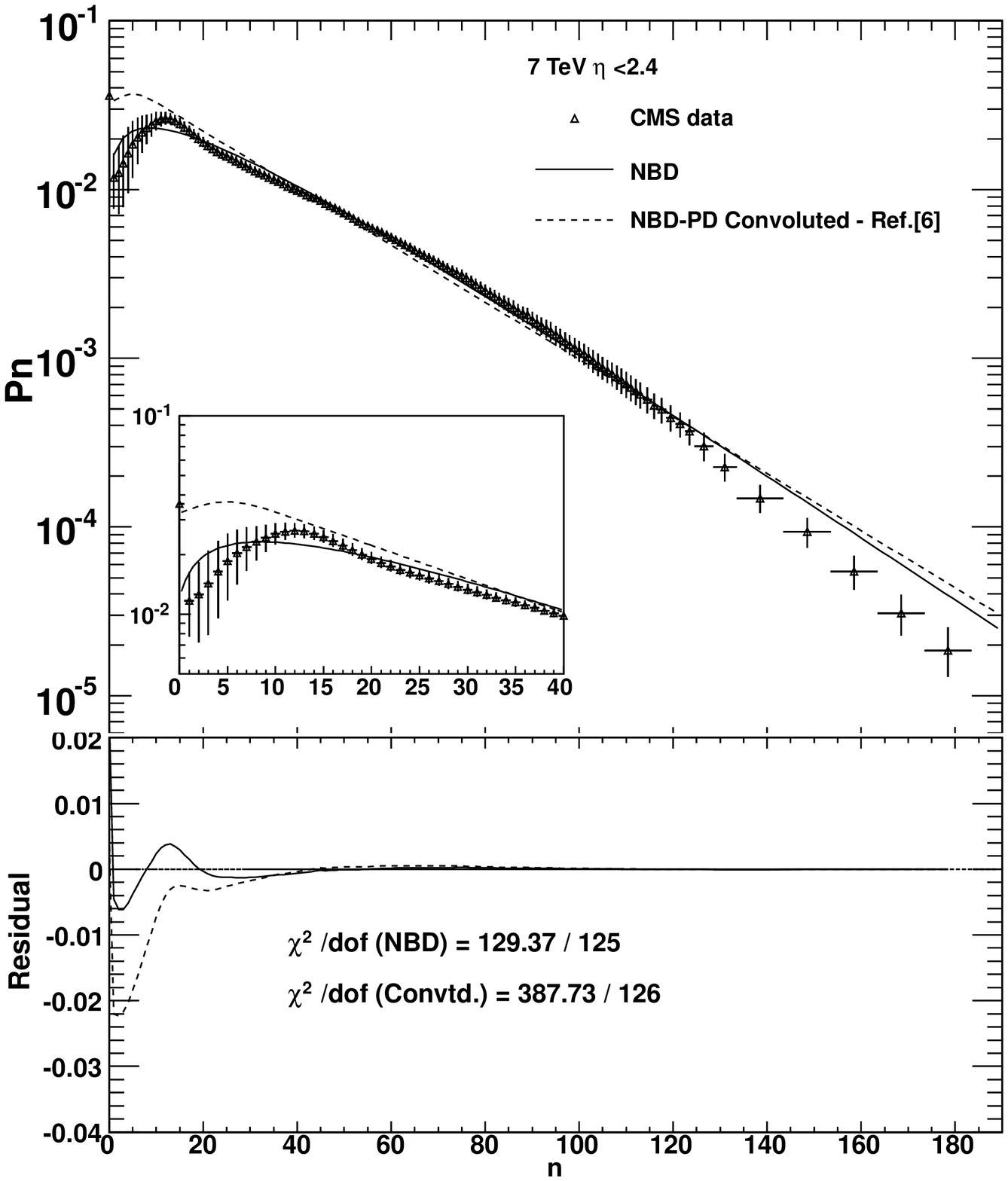}
\caption{Primary charged hadron multiplicity distributions for $|\eta|<$2.4 for $\sqrt {s}$ = 7.0 TeV. The continuous line corresponds to fit to a NBD function and the dotted
line corresponds to the convolution of a NBD and a PD with parameters derived from the two component model as described in Ref.[6]. The lower part of the panel contains 
$\chi ^2$/dof as well as the plots of residual analysis to test goodness of the fits. The error-bars shown with the data-points include both the statistical and the 
systematic uncertainties.}
\label{fig:distribution1} 
\end{figure}
\end{center}
It may be noted that the discussed scaling law is not a characteristic feature of the 
two-component model. The scaling law was developed for information entropy in quantum statistical 
point of view, modifying the entropy scaling \cite{ref19} found to obey experimentally 
measured multiplicity distribution in hadronic interactions for wide range of $\sqrt {s}$, including ISR and SPS data. 
The response of the LHC-data to the scaling law is worth observing. We study the scaling law with the CMS data of $pp$ 
collisions at $\sqrt {s} =$ 0.9, 2.36 and 7 TeV along with the SPS data of $p\bar p$ at $\sqrt {s} =$540 GeV. The data points 
along with respective experimental uncertainties are plotted and are joined with straight lines for a given $\sqrt {s}$ to 
guide the eye. As it is clear from the Fig.~\ref{fig:entropy_scaling}, the data points of different $\sqrt {s}$, do not follow a 
common single curve. Moreover, data-points of different energies follow distinctly separate lines. This observation 
indicates to the deviation of the entropy-scaling at LHC energies.
\begin{center}
\begin{figure}[htbp]
\includegraphics[scale=0.44]{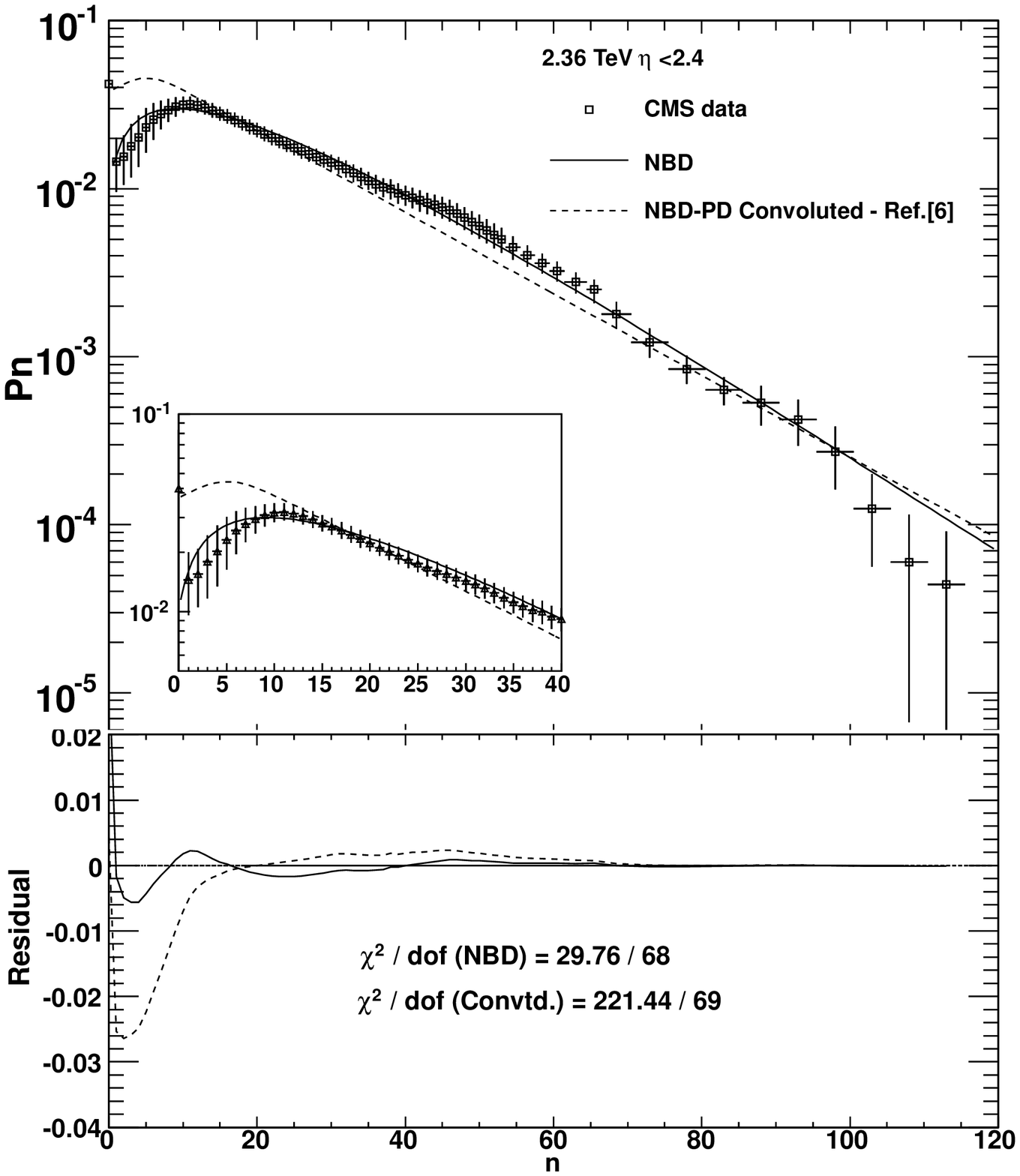}
\caption{Same as Figure - 4 for $\sqrt {s}$ = 2.36 TeV. Lines of different styles correspond to different fits similar to those as described in the caption of Figure -4.
The lower part of the panel contains $\chi ^2$/dof as well as the plots of residual analysis to test goodness of the fits. The error-bars shown with the data-points include 
both the statistical and the systematic uncertainties.}
\label{fig:distribution2} 
\end{figure}
\end{center}
\subsection{Test of agreement between the model and the data.}
\label{}
At this stage, one might be interested to know how well the discussed two-component model describes the measured multiplicity 
distributions. To check the agreement between the model and the data, we chose multiplicity distributions 
in the pseudorapidity interval, $|\eta|<$2.4 for the fact that the change in slopes in distributions have been observed by 
the CMS collaboration for in the same $\eta$-interval with most pronounced structure appearing at $\sqrt {s}$ = 7 TeV and that the motivation of this
work has been the understanding of change in slope of the multiplicity distribution by the two-component model.

\begin{center}
\begin{figure}[htbp]
\includegraphics[scale=0.44]{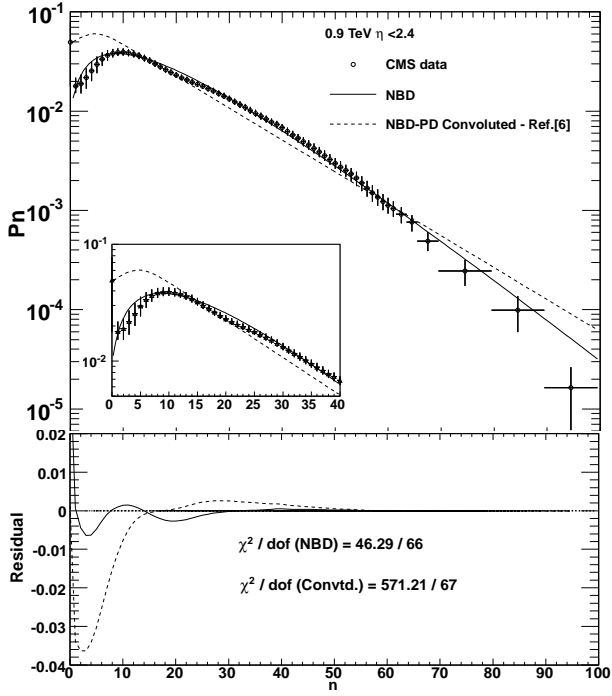}
\caption{Same as Figure - 4 for $\sqrt {s}$ = 0.9 TeV. Lines of different styles correspond to different fits similar to those as described in the caption of Figure -4.
The lower part of the panel contains $\chi ^2$/dof as well as the plots of residual analysis to test goodness of the fits. The error-bars shown with the data-points include 
both the statistical and the systematic uncertainties.}
\label{fig:distribution3} 
\end{figure}
\end{center}

We fit the multiplicity distributions of primary charged hadrons, as measured by the CMS experiment, at energies 
$\sqrt {s}$ = 0.9, 2.36 and 7 TeV in the pseudorapidity interval ($\eta_{c}<$2.4) around 
the mid-$\eta$ with NBD. For clarity in presentation, plots of three energies are given separately in 
Fig.~\ref{fig:distribution1},~\ref{fig:distribution2},~\ref{fig:distribution3}.
The lower panel of each of these figures contains information on goodness of respective fits.
Interestingly, all three distributions, including the one for $\eta_{c}<$2.4 at $\sqrt{s} = 7$ TeV, 
where a prominent change in slope appears, match well with NBD as it is evident from the values of 
the  $\chi ^2$/dof, quoted in respective figures. In the same figures the dotted lines correspond
to the fit with the convolution of a NBD (with k=1) and a PD with parameters extracted from the data 
by the discussed two component model. 
As can be seen by visual inspection of the plots (with zoomed insets) as well as from the corresponding 
plots of residual analysis and the values of $\chi^2$/dof, the multiplicity distributions fit better with single NBD 
than with the convolution of a NBD and a PD in terms of the discussed two component model, indicating poor agreement between 
the model and the LHC data of multiplicity distribution. The residual is defined \cite{ref11} 
as the difference between a data-point and corresponding fit-value.   
\section{Summary and Remarks}
\label{}
We have studied multiplicity distributions of primary charged hadrons produced in 
proton-proton collisions of non-single diffractive class of events at LHC energies, 
in terms of a two-component model of particle production based on quantum statistical 
formalism. Though the behaviour of characteristic parameters of the formalism is found to be
consistent with the LHC data, a scaling law of quantum statistical information entropy, calculated 
from chaotic multiplicity, extracted from the LHC data by the model, is violated.

A test of agreement between the model and the data by fitting the multiplicity
distributions with convolutions of functions as suggested by the model and using values 
of the characteristic parameters evaluated from the data reveals that the model indeed fails 
to describe the data satisfactorily. On the other hand, the commonly used single NBD function fits 
better than the convolution suggested by the model. Our analysis thus shows disagreement 
between the LHC data and the discussed two-component quantum statistical model in its present form.

\end{document}